\documentclass{patmorin}
\usepackage[utf8]{inputenc}
\usepackage{amsthm,amsmath,graphicx,wrapfig}
\usepackage[noend]{algorithmic}
\usepackage{pat}
\usepackage{tikz,gnuplot-lua-tikz}

\newcommand{\eps}{\varepsilon}

\title{\MakeUppercase{Top-Down Skiplists}}
\author{Luis Barba%
     \thanks{School of Computer Science, Carleton University
             and Département d'Informatique, 
             Université Libre de Bruxelles,
             \email{lbarbafl@ulb.ac.be}}\enspace
     and Pat Morin%
     \thanks{School of Computer Science, Carleton University,
             \email{morin@scs.carleton.ca}}}

\begin{document}
\begin{titlepage}
\maketitle

\begin{abstract}
  We describe todolists (top-down skiplists), a variant of skiplists
  (Pugh 1990) that can execute searches using at most $\log_{2-\eps} n +
  O(1)$ binary comparisons per search and that have amortized update time
  $O(\eps^{-1}\log n)$. A variant of todolists, called working-todolists,
  can execute a search for any element $x$ using $\log_{2-\eps} w(x)
  + o(\log w(x))$ binary comparisons and have amortized search time
  $O(\eps^{-1}\log w(w))$. Here, $w(x)$ is the ``working-set number'' of
  $x$. No previous data structure is known to achieve a bound better
  than $4\log_2 w(x)$ comparisons. We show through experiments that,
  if implemented carefully, todolists are comparable to other common
  dictionary implementations in terms of insertion times and outperform
  them in terms of search times.
\end{abstract}

\end{titlepage}

\section{Introduction}

Comparison-based dictionaries supporting the three \emph{basic
operations} insert, delete and search represent \emph{the} classic
data-structuring problem in computer science.  Data structures that
support each of these operations in $O(\log n)$ time have been known
since the introduction of AVL trees more than half a century ago
\cite{adelson-velskii.landis:algorithm}.  Since then, many competing
implementations of dictionaries have been proposed, including
red-black trees \cite{guibas.sedgewick:dichromatic}, splay trees
\cite{sleator.tarjan:self-adjusting}, general-balanced/scapegoat
trees \cite{andersson:general,galperin.rivest:scapegoat},
randomized binary search trees \cite{martinez:randomized},
energy-balanced trees \cite{goodrich:competitive}, Cartesian trees/treaps
\cite{aragon.seidel:randomized,vuillemin:unifying}, skip lists
\cite{pugh:skip}, jump lists \cite{bronnimann.cazals.ea:randomized},
and others.  Most major programming environments include one or more
$O(\log n)$ time dictionary data structures in their standard library,
including Java's \texttt{TreeMap} and \texttt{TreeSet}, the C++ STL's
\texttt{set}, and Python's \texttt{OrderedDict}.

In short, comparison-based dictionaries are so important that any
new ideas or insights about them are worth exploring.  In this paper,
we introduce the todolist (\boldx{to}p-\boldx{do}wn skip\boldx{list}),
a dictionary that is parameterized by a parameter $\eps\in(0,1)$, 
that can execute searches using at most $\log_{2-\eps} n + O(1)$ binary
comparisons per search, and that has amortized update time $O(\eps^{-1}\log
n)$.  (Note that $\log_{2-\eps} n \le (1+\eps)\log n$ for $\eps < 1/4$.)

As a theoretical result todolists are nothing to get excited about; there
already exist comparison-based dictionaries with $O(\log n)$ time for all
operations that perform at most $\lceil\log n\rceil+1$ comparisons per
operation \cite{andersson.lai:fast,fagerberg:binary}. (Here, and throughout,
$\log n=\log_2 n$ denotes the binary logarithm of $n$. However, todolists
are based on a new idea---top-down partial rebuilding of skiplists---and
our experimental results show that a careful implementation of todolists
can execute searches faster than existing popular data structures.

In particular, todolists outperform (again, in terms of searches) Guibas
and Sedgewick's red-black trees \cite{guibas.sedgewick:dichromatic} which are
easily the most common implementation of comparison-based dictionaries
found in programming libraries. This is no small feat since, in the
setting we studied, the average depth of a node in a red-black tree
seems to be $\log n - O(1)$ \cite{sedgewick:left-leaning}.

As a more substantial theoretical contribution, we show that a variant of todolists, called working-todolists, is able to search for
an element $x$ using $\log_{2-\eps} w(x)+o(\log w(x))$ comparisons
in $O(\eps^{-1}\log w(x))$ amortized time.  Here, $w(x)$---the
working set number of $x$---is loosely defined as the number of
distinct items accessed since the last time $x$ was accessed (see
\secref{working-todolist} for a precise definition.)  Previous data
structures with some variant of this \emph{working-set property}
include splay trees \cite{sleator.tarjan:self-adjusting}, Iacono's
working-set structure \cite{iacono:alternatives,badoiu.cole.ea:unified},
deterministic self-adjusting skiplists \cite{bose.douieb.ea:dynamic},
layered working-set trees \cite{bose.douieb.ea:layered}, and skip-splay
trees \cite{derryberry.sleator:skip-splay}.  However, even the most
efficient of these can only be shown to perform at most $4\log w(x)$
comparisons during a search for $x$.

\section{TodoLists}
\seclabel{todolist}

A \emph{todolist} for the values $x_1<x_2<\cdots<x_n$ consists of a
nested sequence of $h+1$ sorted singly-linked lists, $L_0,\ldots,L_h$,
having the following properties:\footnote{Here and throughout, we use set
notations like $|\cdot|$, and $\subseteq$ on the lists $L_0,\ldots,L_h$,
with the obvious interpretations.}

\begin{enumerate}
  \item $|L_0| \le 1$.
  \item $L_i\subseteq L_{i+1}$ for each $i\in\{0,\ldots,h-1\}$.
  \item For each $i\in\{1,\ldots,h\}$ and each pair $x,y$ of consecutive
        elements in $L_i$, at least one of $x$ or $y$ is in $L_{i-1}$.
  \item $L_h$ contains $x_1,\ldots,x_n$.
\end{enumerate}

The value of $h$ is at least $\lceil \log_{2-\varepsilon} n\rceil$ and
at most $\lceil \log_{2-\varepsilon} n\rceil+1$.  The \emph{height} of
a value, $x$, in a todolist is the number of lists in which $x$ appears.

We will assume that the head of each list $L_i$ is a \emph{sentinel} node that does not contain any data. (See \figref{todolist}.)  We will also
assume that, given a pointer to the node containing $x_j$ in $L_i$, it is
possible to find, in constant time, the occurrence of $x_j$ in $L_{i+1}$.
This can be achieved by maintaining an extra pointer or by maintaining
all occurrences of $x_j$ in an array. (See \secref{implementation} for
a detailed description.)

\begin{figure}
  \centering{\includegraphics{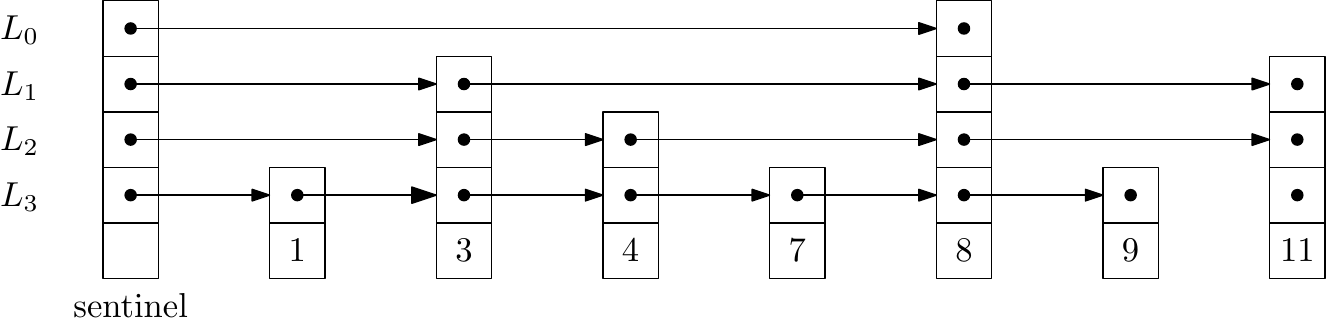}}
  \caption{An example of a todolist containing $1,3,4,7,8,9,11$.}
  \figlabel{todolist}
\end{figure}

\subsection{Searching}

Searching for a value, $x$, in a todolist is simple. In particular, we
can find the node, $u$, in $L_h$ that contains the largest value that
is less than $x$. If $L_h$ has no value less than $x$ then the search
finds the sentinel in $L_h$.  We call the node $u$ the \emph{predecessor}
of $x$ in $L_h$.

Starting at the sentinel in $L_0$, one comparison (with the at most one
element of $L_0$) is sufficient to determine the predecessor, $u_0$ of $x$
in $L_0$. (This follows from Property~1.)  Moving down to the occurrence
of $u_0$ in $L_1$, one additional comparison is sufficient to determine
the predecessor, $u_1$ of $x$ in $L_1$. (This follows from Property~3.)
In general, once we know the predecessor of $x$ in $L_i$ we can determine
the predecessor of $x$ in $L_{i+1}$ using one additional comparison. Thus,
the total number of comparisons needed to find the predecessor of $x$
in $L_h$ is only $h+1$.

\vspace{1ex}
\noindent{$\textsc{FindPredecessor}(x)$}
\begin{algorithmic}
  \STATE{$u_0\gets \mathtt{sentinel}_0$}
  \FOR{$i=0,\ldots,h$}
    \IF{$\mathrm{next}(u_i)\neq \mathbf{nil}$ and $\mathrm{key}(\mathrm{next}(u_i)) < x$}
      \STATE{$u_i\gets\mathrm{next}(u_i)$}
    \ENDIF
    \STATE{$u_{i+1}\gets\mathrm{down}(u_i)$}
  \ENDFOR
  \RETURN{$u_h$}
\end{algorithmic}

\subsection{Adding}
\seclabel{insertion}\seclabel{adding}

Adding a new element, $x$, to a todolist is done by searching for it
using the algorithm outlined above and then splicing $x$ into each of
the lists $L_0,\ldots,L_h$.  This splicing is easily done in constant
time per list, since the new nodes containing $x$ appear after the nodes
$u_0,\ldots,u_h$.  At this point, all of the Properties~2--4 are satisfied,
but Property~1 may be violated since there may be two values in $L_0$.

If there are two values in $L_0$, then we restore Property~1 with the
following \emph{partial rebuilding} operation: We find the smallest index
$i$ such that $|L_i|\le (2-\eps)^i$; such an index always exists since
$n=|L_h|\le(2-\eps)^h$.  We then rebuild the lists $L_{0},\ldots,L_{i-1}$
in a bottom up fashion; $L_{i-1}$ gets every second element from $L_i$
(starting with the second), $L_{i-2}$ gets every second element from
$L_{i-1}$, and so on down to $L_0$.

Since we take every other element from $L_i$ starting with the second element,
after rebuilding we obtain:
\[
   |L_{i-1}| = \lfloor |L_i|/2 \rfloor \le |L_i|/2
\]
and, repeating this reasoning for $L_{i-2}, L_{i-3},\ldots, L_0$, we see that, after rebuilding,
\[
   |L_{0}| \le |L_i|/2^i \le (2-\eps)^i/2^i < 1 \enspace .
\]
Thus, after this rebuilding, $|L_0|=0$, Property~1 is restored and the
rebuilding, by construction, produces lists satisfying Properties~2--4.

To study the amortized cost of adding an element, we
can use the potential method with the potential function
\[
    \Phi(L_0,\ldots,L_h)=C\sum_{i=0}^h|L_i| \enspace .
\]
Adding $x$ to each of $L_0,\ldots,L_h$ increases this potential by
$C(h+1)=O(C\log n)$.  Rebuilding, if it occurs, takes $O(|L_i|)=O((2-\eps)^i)$
time, but causes a change in potential of at least
\begin{align*}
     \Delta\Phi & = C\sum_{j=0}^i\left(|L_j| - (2-\eps)^j\right) \\
     	& = C\sum_{j=0}^i\left(|L_i|/2^{i-j} - (2-\eps)^j\right) \\ 
          & \le C\sum_{j=0}^{i-1}\left((2-\eps)^i/2^{i-j} - (2-\eps)^j\right) \\
          & \le C\left((2-\eps)^i - \sum_{j=0}^{i-1}(2-\eps)^j\right) \\
          & = C\left((2-\eps)^i - \frac{(2-\eps)^i-(2-\eps)}{1-\eps}\right) \\
          & < C\left((2-\eps)^i - (1+\eps)\left((2-\eps)^i-(2-\eps)\right)\right)
           & \text{(since $1/(1-\eps)>1+\eps$)} \\
          & = -C\eps(2-\eps)^i + O(C) \\
\end{align*}
Therefore, by setting $C=c/\eps$ for a sufficiently large constant,
$c$, the decrease in potential is greater than the cost of rebuilding.
We conclude that the amortized cost of adding an element $x$ is $O(C\log
n)=O(\eps^{-1}\log n)$.

\subsection{Deleting}

Since we already have an efficient method of partial rebuilding, we
can use it for deletion as well. To delete an element $x$, we delete
it in the obvious way, by searching for it and then splicing it out
of the lists $L_i,\ldots,L_h$ in which it appears.  At this point,
Properties~1, 2, and 4 hold, but Property~3 may be violated in any
subset of the lists $L_i,\ldots,L_h$.  Luckily, all of these violations
can be fixed by taking $x$'s successor in $L_h$ and splicing it into
each of $L_0,\ldots,L_{h-1}$.\footnote{If $x$ has no successor in
$L_h$---because it is the largest value in the todolist---then deleting
$x$ will not introduce any violations of Property~3.}  Thus, the second
part of the deletion operation is like the second part of the insertion
operation.  Like the insertion operation, this may violate Property~1
and trigger a partial rebuilding operation.  The same analysis used to
study insertion shows that deletion has the same amortized running time
of $O(\eps^{-1}\log n)$.

\subsection{Tidying Up}

Aside from the partial rebuilding caused by insertions and deletions,
there are also some \emph{global rebuilding} operations that are sometimes
triggered:
\begin{enumerate}
\item If an insertion causes $n$ to exceed $\lceil(2-\eps)^h\rceil$, then
we increment the value of $h$ to $h'=h+1$ and rebuild $L_0,\ldots,L_{h'}$
from scratch, starting by moving $L_h$ into $L_{h'}$ and then performing
a partial rebuilding operation on $L_{0},\ldots,L_{h'-1}$.
\item If an insertion or deletion causes $\sum_{i=1}^n |L_i|$ to exceed $cn$ for some threshold constant $c>2$, then we perform a partial rebuilding to rebuild $L_{0},\ldots,L_{h-1}$.
\item If a deletion causes $n$ to be less than $\lceil(2-\eps)^{h-2}\rceil$ then we decrement the value of $h$ to be $h'=h-1$, move $L_h$ to $L_{h'}$ and then perform a partial rebuilding operation on $L_{0},\ldots,L_{h'-1}$. 
\end{enumerate}

A standard amortization argument shows that the first and third type
of global rebuilding contribute only $O(1)$ to the amortized cost of
each insertion and deletion, respectively.  The same potential function
argument used to study insertion and deletion works to show that the
second type of global rebuilding contributes only $O(\log n)$ to the
amortized cost of each insertion or deletion (note that this second
type of global rebuilding is only required to ensure that the size of
the data structure remains in $O(n)$).

This completes the proof of our first theorem:
\begin{thm}\thmlabel{todolist}
For any $\eps >0$, a todolist supports the operations of inserting,
deleting, and searching using at most $\log_{2-\eps} n + O(1)$ comparisons
per operation.  Starting with an empty todolist and performing any
sequence of $N$ add and remove operations takes $O(\eps^{-1}N\log
N)$ time.
\end{thm}

\section{Working-TodoLists}
\seclabel{working-todolist}

Next, we present a new theoretical result that is achieved using a
variant of the todolist that we call a working-todolist.  First, though, we need
some definitions.  Let $a_1,\ldots,a_m$ be a sequence whose elements come
from the set $\{1,\ldots,n\}$.  We call such a sequence an \emph{access
sequence}. For any $x\in\{1,\ldots,n\}$, the \emph{last-occurrence},
$\ell_t(x)$, of $x$ at time $t$ is defined as
\[
   \ell_t(x)=\max\{j\in{1,\ldots,t-1}: a_{j} = x\} \enspace .
\]
Note that $\ell_t(x)$ is undefined if $x$ does not appear in
$a_1,\ldots,a_{t-1}$.  The \emph{working-set number}, $w_t(x)$, of $x$
at time $t$ is
\[
    w_t(x) = \begin{cases}
               n & \text{if $\ell_t(x)$ is undefined} \\
               |\{a_{\ell_t(x)},\ldots,a_{t-1}\}| & \text{otherwise.}
             \end{cases}
\]
In words, if we think of $t$ as the current time, then $w_t(x)$ is the
number of distinct values in the access sequence since the most recent
access to $x$.

In this section, we describe the working-todolist data structure, which
stores $\{1,\ldots,n\}$ and, for any access sequence $a_1,\ldots,a_m$,
can execute searches for $a_1,\ldots,a_m$ so that the search for $a_t$
performs at most $(1+o(1))\log_{2-\eps} w_t(a_t)$ comparisons and takes
$O(\eps^{-1}\log w_t(a_t))$ amortized time.

From this point onward we will drop the time subscript, $t$, on $w_t$
and assume that $w(x)$ refers to the working set number of $x$ at the
current point in time (given the sequence of searches performed so far).
The working-todolist is a special kind of todolist that weakens Property~1
and adds an additional Property~5:

\begin{enumerate}
\item $|L_0|\le \eps^{-1}+1$.
\setcounter{enumi}{4}
\item For each $i\in\{0,\ldots,h\}$, $L_i$ contains all values $x$ such that $w(x)\le (2-\eps)^i$.
\end{enumerate}

For keeping track of working set numbers, a working-todolist also stores a
doubly-linked list, $Q$, that contains the values $\{1,\ldots,n\}$
ordered by their current working set numbers.  The node that contains $x$
in this list is cross-linked (or associated in some other way) with the
appearances of $x$ in $L_0,\ldots,L_h$.

\subsection{Searching}
\seclabel{todolist-search}

Searching in a working-todolist is similar to a searching in a todolist.
The main
difference is that Property~5 guarantees that the working-todolist will reach
a list, $L_i$, that contains $x$ for some $i\le\log_{2-\eps} w(x)$.
If ternary comparisons are available, then this is detected at the
first such index $i$.  If only binary comparisons are available, then
the search algorithm is modified slightly so that, at each list $L_i$
where $i$ is a perfect square, an extra comparison is done to test if
the successor of $x$ in $L_i$ contains the value $x$.  This modification
ensures that, if $x$ appears first in $L_i$, then it is found by the
time we reach the list $L_{i'}$ for
\[
     i'=i+\left\lceil 2\sqrt{i}\right\rceil + 1 = \log_{2-\eps} w(x) + O(\sqrt{\log w(x)}) \enspace .
\]

Once we find $x$ in some list $L_{i'}$, we move it to the front of $Q$;
this takes only constant time since the node containing $x$ in $L_{i'}$
is cross-linked with $x$'s occurrence in $Q$.  Next, we insert $x$ into
$L_0,\ldots,L_{i'-1}$.  As with insertion in a todolist, this takes only
constant time for each list $L_j$, since we have already seen the predecessor
of $x$ in $L_j$ while searching for $x$.  
At this point, Properties 2--5 are ensured and the ordering of $Q$ is
correct.  

All that remains is to restore Property~1, which is now violated
since $L_0$ contains $x$, for which $w(x)=1$, and the value $y$ such
that $w(y)=2$.  Again, this is corrected using partial rebuilding,
but the rebuilding is somewhat more delicate.  We find the first index
$i$ such that $|L_i|\le (2-\eps/2)^i$.  Next, we traverse the first
$(2-\eps)^{i-1}$ nodes of $Q$ and label them with their position in $Q$.
Since $Q$ is ordered by working-set number, this means that the label
at a node of $Q$ that contains the value $z$ is at most $w(z)$.

At this point, we are ready to rebuild the lists $L_0,\ldots,L_{i-1}$. To
build $L_{j-1}$ we walk through $L_j$ and take any value whose label
(in $Q$) is defined and is at most $(2-\eps)^j$ as well as every
``second value'' as needed to ensure that Property~3 holds.  Finally,
once all the lists $L_0,\ldots,L_j$ are rebuilt, we walk through the first
$(2-\eps)^{i-1}$ nodes of $Q$ and remove their labels so that these labels
are not incorrectly used during subsequent partial rebuilding operations.

\subsection{Analysis}

We have already argued that we find a node containing $x$ in some list
$L_i$ with $i\in \log_{2-\eps} w(x) + O(\sqrt{\log w(x)})$ and that this
takes $O(\log w(x))$ time.  The number of comparisons needed to reach
this stage is
\[
     \log_{2-\eps} w(x) + O\left(\eps^{-1} + \sqrt{\log w(x)}\right) \enspace .
\]
The $O(\eps^{-1})$ term is the cost of searching in $L_0$ and the
$O(\sqrt{\log w(x)})$ term accounts for one comparison at each of the lists
$L_{\lceil\log_{2-\eps} w(x)\rceil},\ldots,L_i$ as well as the extra
comparison performed in each of the lists $L_j$ where $j\in\{0,\ldots,i\}$
is a perfect square.

After finding $x$ in $L_i$, the algorithm then updates
$L_0,\ldots,L_{i-1}$ in such a way that Properties~2--5 are maintained.
All that remains is to show that Property~1 is restored by the partial
rebuilding operation and to study the amortized cost of this partial
rebuilding.  We accomplish both these goals by studying the sizes of the lists
$L_0,\ldots,L_i$ after rebuilding.

Let $n_i=|L_i|$ and recall that $n_i\le (2-\eps/2)^i$. Then,
the number of elements that make it from $L_i$ into $L_{i-1}$ is 
\[  |L_{i-1}| \le (2-\eps)^{i-1} + n_i/2 \enspace , \]
and the number of elements that make it into $L_{i-2}$ is
\begin{align*}
   |L_{i-2}| & \le (2-\eps)^{i-2} + |L_{i-1}|/2 \\
     & \le (2-\eps)^{i-2} + (2-\eps)^{i-1}/2 + n_i/4 \enspace . 
\end{align*}
More generally, the number of elements that make it into $L_j$ for any $j\in \{0,\ldots,i-1\}$ is at most
\begin{align*}
    |L_j| & \le (2-\eps)^{j} \cdot \sum_{k=0}^{i-j-1}\left(\frac{2-\eps}{2}\right)^k  + n_i/2^{i-j} \\
       & \le (2-\eps)^j/\eps + n_i/2^{i-j} \enspace .
\end{align*}
In particular
\[
    |L_0| \le \eps^{-1} + n_i/2^i \le \eps^{-1} + 1\enspace .
\]
Therefore, Property~1 is satisfied.

To study the amortized cost of searching for $x$, we use the same
potential function argument as in \secref{todolist}.  The change in the sizes of the lists is then 
\begin{align*}
  \Delta\Phi/C & \le \sum_{j=0}^{i-1}\left((2-\eps)^j/\eps + n_i/2^{i-j} - (2-\eps/2)^j\right) \\
  & \le O((2-\eps)^i/\eps) + n_i - \sum_{j=0}^{i-1}(2-\eps/2)^j \\
  & = O((2-\eps)^i/\eps) + n_i - \frac{(2-\eps/2)^i}{1-\eps/2} + O(1) \\
  & \le O((2-\eps)^i/\eps) + n_i - (1+\eps/2)((2-\eps/2)^i) \\
  & \le O((2-\eps)^i/\eps) + n_i - (1+\eps/2)n_i \\
  & \le O((2-\eps)^i/\eps) - (\eps/2)n_i \\
  & = - \Omega(\eps n_i) & \text{(since $n_i \ge n_{i-1} \ge (2-\eps/2)^{i-1}$)}
\end{align*}
Since the cost of rebuilding $L_0,\ldots,L_{i-1}$ is $O(n_i)$, this implies that the amortized cost of accessing $x$ is $O(\eps^{-1}\log w(x))$.

\section{Implementation Issues}
\seclabel{implementation}

As a first attempt, one might try to implement a todolist exactly
as described in \secref{todolist}, with each list $L_i$ being a
separate singly linked list in which each node has a down pointer to
the corresponding node in $L_{i+1}$.  However, past experience with
skiplists suggests (and preliminary experiments confirms) that this is
neither space-efficient nor fast (see the class \texttt{LinkedTodoList}
in the source code). Instead, we use an implementation idea that appears in
Pugh's original paper on skiplists~\cite{pugh:skip}.

\subsection{Nodes as Arrays}

A better implementation uses one structure for each data item, $x$,
and this structure includes an array of pointers.  If $x$ appears in
$L_{i},\ldots,L_h$, then this array has length $h-i+1$ so that it can
store the \texttt{next} pointers for the occurrence of $x$ in each of
these lists.  

One complication occurs in implementing arrays of next pointers in a
todolist that is not present in a skiplist.  During a partial rebuilding
operation, the heights of elements in a todolist change, which means
that their arrays need to be reallocated.  The cost of reallocating and
initializing an array is proportional to the length of the array. However,
the amortization argument used in \secref{insertion}, requires that the
cost of increasing the height of an element when rebuilding level $i$
is proportional to the increase in height; promoting an element from
level $i$ to level $i-c$ should take $O(c)$ time, not $O(h - i + c)$ time.

The work-around for this problem is to use a standard
doubling trick used to implement dynamic arrays (c.f.,
Morin~\cite[Section~2.1.2]{morin:open}). When a new array for a node
is allocated to hold $r$ values, its size is set to $r'=2^{\lceil\log
r\rceil}$.  Later, if the height of the node increases, to $r+c$ during
a partial rebuilding operation, the array only needs to be reallocated
if $r+c > r'$.  Using this trick, the amortized
cost of increasing the height of the node is $O(c)$.  This trick does not
increase the number of pointers in the structure by more than factor of 2.

Our initial implementation did exactly this, and performed well-enough
to execute searches faster than standard skiplists but was still bested
by most forms of binary search trees.  This was despite the fact that
the code for searching was dead-simple, and by decreasing $\eps$ we
could reduce the height, $h$, (and hence the number of comparisons)
to less than was being performed by these search trees.

\subsection{The Problem With Skiplists}

After some reflection, the reason for the disappointing performance of
searches in todolists (and to a greater extent, in skiplists) became
apparent. It is due to the fact that accessing a node by following a
pointer that causes a CPU cache miss is more expensive than performing
a comparison.

The search path in a todolist has length equal to the number of
comparisons performed.  However, the set of nodes in the todolist
that are dereferenced during a search includes nodes not on the
search path. Indeed, when the outcome of a comparison of the form
$\mathrm{key}(\mathrm{next}(u)) < x$ is false, the search path proceeds
to $\mathrm{down}(u)$ and the node $\mathrm{next}(u)$ does not appear on
the search path.

\subsection{The Solution}

Luckily, there is a fairly easy remedy, though it does use more space.
We implement the todolist so that each node $u$ in a list $L_i$ stores an
additional key, $\mathrm{keynext}(u)$, that is the key associated with
the node $\mathrm{next}(u)$.  This means that determining the next node
to visit after $u$ can be done using the key, $\mathrm{keynext}(u)$, stored
at node $u$ rather than having to dereference $\mathrm{next}(u)$. The
resulting structure is illustrated in \figref{packed-in}.

\begin{figure}
   \centerline{\includegraphics{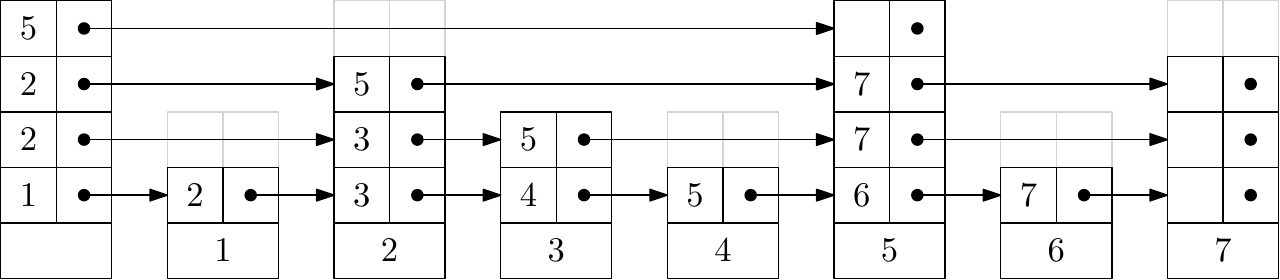}}
   \caption{The memory layout of an efficient todolist implementation.}
   \figlabel{packed-in}
\end{figure}

With this modification, todolists achieve faster search times---even with
fairly large values of $\eps$---than binary search tree structures. Indeed,
retrofitting this idea into the skiplist implementation improves its
performance considerably so that it outperforms some (but not all)
of the tree-based structures.

\subsection{Experiments}
\seclabel{experiments}

To test the performance of todolists, we implemented them and tested them
against other comparison-based dictionaries that are popular, either in
practice (red-black trees) and/or in textbooks (scapegoat trees, treaps,
and skiplists).  The implementation of all data structures was done in
C++ and all the code was written by the second author.\footnote{The
implementations of all but todolists were adapted from the second
author's textbook \cite{morin:open}.}  To ensure that this code is
comparable to so-called industrial strength C++ code, the tests also
include the C++ Standard Template Library \texttt{set} class that comes
as part of \texttt{libsdc++}.  This \texttt{set} class is implemented
as a red-black tree and performed indistinguishably from our red-black
tree implementation.

The code used to generate all the test data in this section is available
for download at github.\footnote{The source code is available at
\url{https://github.com/patmorin/todolist}. The final version of this
paper will provide a digital object identifier (DOI) that provides a
link to the permanent fixed version of the source code that was used to
generate all data in the final paper.}

The experimental data in this section was obtained from the program
\texttt{main.cpp} that can be found in the accompanying source code.
This program was compiled using the command line: \texttt{g++ -std=c++11 -Wall -O4 -o main main.cpp}. The compiler was the \texttt{gcc} compiler, version
4.8.2 that ships with the Ubuntu 14.04 Linux distribution.  Tests were
run on a desktop computer having 16GB DDR3 1600MHz memory and a Intel
Core i5-4670K processor with 6MB L3 cache and running at 3.4GHz.

\subsubsection{Varying $\eps$}

\Figref{epsilon} shows the results of varying the value of $\eps$ from
$0.02$ to $0.68$ in increments of $0.01$. In this figure, $n=10^6$
random integers in the set $\{0, 5,\ldots,5(n-1)\}$ were chosen
(with replacement) and inserted into a todolist. Since dictionaries
discard duplicates, the resulting todolist contained $906,086$ values.
This todolist was then searched $m=5n$ times with random integers chosen,
with replacement, from $\{-2,\ldots,5n+3\}$.  This figure illustrates
that todolists do behave roughly as \thmref{todolist} predicts.
Insertion time increases roughly proportionally to $1/\eps$ and search
times seem to be of the form $c(d+\eps)$ for some constant $c$ and $d$
(though there is certainly lots of noise in the search times).

In terms of implementation guidance, this figure suggests that values
of $\eps$ below $0.1$ are hard to justify.  The improvement in search
time does not offset the increase in insertion time.  At the same
time, values of $\eps$ greater than $0.35$ do not seem to be of much
use either since they increase the search time and don't decrease the
insertion time significantly. At some point beyond this---at around
$\eps=0.45$---increasing $\eps$ increases both the insertion time and
the search time (since every insertion starts with a search).

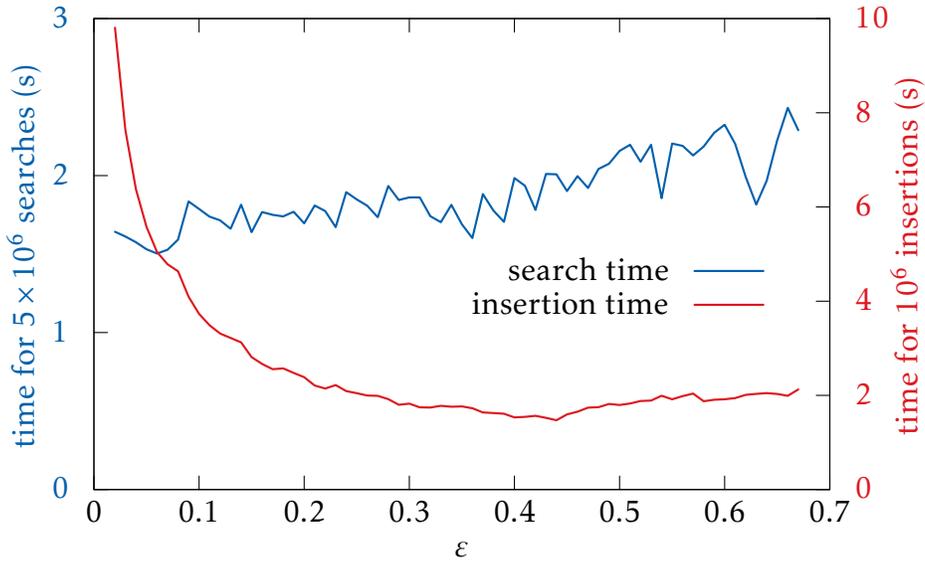
\begin{figure}
  \centering{\begin{tikzpicture}[gnuplot]
\path (0.000,0.000) rectangle (12.700,7.620);
\gpcolor{color=gp lt color border}
\gpsetlinetype{gp lt border}
\gpsetlinewidth{1.00}
\draw[gp path] (1.136,0.985)--(1.316,0.985);
\gpcolor{rgb color={0.000,0.376,0.678}}
\node[gp node right] at (0.952,0.985) { 0};
\gpcolor{color=gp lt color border}
\draw[gp path] (1.136,3.074)--(1.316,3.074);
\gpcolor{rgb color={0.000,0.376,0.678}}
\node[gp node right] at (0.952,3.074) { 1};
\gpcolor{color=gp lt color border}
\draw[gp path] (1.136,5.162)--(1.316,5.162);
\gpcolor{rgb color={0.000,0.376,0.678}}
\node[gp node right] at (0.952,5.162) { 2};
\gpcolor{color=gp lt color border}
\draw[gp path] (1.136,7.251)--(1.316,7.251);
\gpcolor{rgb color={0.000,0.376,0.678}}
\node[gp node right] at (0.952,7.251) { 3};
\gpcolor{color=gp lt color border}
\draw[gp path] (1.136,0.985)--(1.136,1.165);
\draw[gp path] (1.136,7.251)--(1.136,7.071);
\node[gp node center] at (1.136,0.677) { 0};
\draw[gp path] (2.534,0.985)--(2.534,1.165);
\draw[gp path] (2.534,7.251)--(2.534,7.071);
\node[gp node center] at (2.534,0.677) { 0.1};
\draw[gp path] (3.931,0.985)--(3.931,1.165);
\draw[gp path] (3.931,7.251)--(3.931,7.071);
\node[gp node center] at (3.931,0.677) { 0.2};
\draw[gp path] (5.329,0.985)--(5.329,1.165);
\draw[gp path] (5.329,7.251)--(5.329,7.071);
\node[gp node center] at (5.329,0.677) { 0.3};
\draw[gp path] (6.726,0.985)--(6.726,1.165);
\draw[gp path] (6.726,7.251)--(6.726,7.071);
\node[gp node center] at (6.726,0.677) { 0.4};
\draw[gp path] (8.124,0.985)--(8.124,1.165);
\draw[gp path] (8.124,7.251)--(8.124,7.071);
\node[gp node center] at (8.124,0.677) { 0.5};
\draw[gp path] (9.521,0.985)--(9.521,1.165);
\draw[gp path] (9.521,7.251)--(9.521,7.071);
\node[gp node center] at (9.521,0.677) { 0.6};
\draw[gp path] (10.919,0.985)--(10.919,1.165);
\draw[gp path] (10.919,7.251)--(10.919,7.071);
\node[gp node center] at (10.919,0.677) { 0.7};
\draw[gp path] (10.919,0.985)--(10.739,0.985);
\gpcolor{rgb color={0.867,0.094,0.122}}
\node[gp node left] at (11.103,0.985) { 0};
\gpcolor{color=gp lt color border}
\draw[gp path] (10.919,2.238)--(10.739,2.238);
\gpcolor{rgb color={0.867,0.094,0.122}}
\node[gp node left] at (11.103,2.238) { 2};
\gpcolor{color=gp lt color border}
\draw[gp path] (10.919,3.491)--(10.739,3.491);
\gpcolor{rgb color={0.867,0.094,0.122}}
\node[gp node left] at (11.103,3.491) { 4};
\gpcolor{color=gp lt color border}
\draw[gp path] (10.919,4.745)--(10.739,4.745);
\gpcolor{rgb color={0.867,0.094,0.122}}
\node[gp node left] at (11.103,4.745) { 6};
\gpcolor{color=gp lt color border}
\draw[gp path] (10.919,5.998)--(10.739,5.998);
\gpcolor{rgb color={0.867,0.094,0.122}}
\node[gp node left] at (11.103,5.998) { 8};
\gpcolor{color=gp lt color border}
\draw[gp path] (10.919,7.251)--(10.739,7.251);
\gpcolor{rgb color={0.867,0.094,0.122}}
\node[gp node left] at (11.103,7.251) { 10};
\gpcolor{color=gp lt color border}
\draw[gp path] (1.136,7.251)--(1.136,0.985)--(10.919,0.985)--(10.919,7.251)--cycle;
\gpcolor{rgb color={0.000,0.376,0.678}}
\node[gp node center,rotate=-270] at (0.246,4.118) {time for $5\times 10^6$ searches (s)};
\gpcolor{rgb color={0.867,0.094,0.122}}
\node[gp node center,rotate=-270] at (11.992,4.118) {time for $10^6$ insertions (s)};
\gpcolor{color=gp lt color border}
\node[gp node center] at (6.027,0.215) {$\varepsilon$};
\node[gp node right] at (8.936,3.893) {search time};
\gpcolor{rgb color={0.000,0.376,0.678}}
\gpsetlinetype{gp lt plot 0}
\gpsetlinewidth{2.00}
\draw[gp path] (9.120,3.893)--(10.036,3.893);
\draw[gp path] (10.500,5.766)--(10.360,6.063)--(10.220,5.633)--(10.080,5.094)--(9.941,4.777)%
  --(9.801,5.144)--(9.661,5.584)--(9.521,5.839)--(9.382,5.732)--(9.242,5.549)--(9.102,5.430)%
  --(8.962,5.558)--(8.823,5.588)--(8.683,4.862)--(8.543,5.572)--(8.403,5.346)--(8.264,5.572)%
  --(8.124,5.490)--(7.984,5.320)--(7.844,5.250)--(7.705,4.997)--(7.565,5.155)--(7.425,4.956)%
  --(7.285,5.179)--(7.146,5.184)--(7.006,4.705)--(6.866,5.027)--(6.726,5.129)--(6.587,4.546)%
  --(6.447,4.697)--(6.307,4.916)--(6.167,4.331)--(6.028,4.514)--(5.888,4.773)--(5.748,4.542)%
  --(5.608,4.622)--(5.468,4.872)--(5.329,4.872)--(5.189,4.837)--(5.049,5.024)--(4.909,4.608)%
  --(4.770,4.761)--(4.630,4.845)--(4.490,4.942)--(4.350,4.476)--(4.211,4.690)--(4.071,4.764)%
  --(3.931,4.527)--(3.791,4.681)--(3.652,4.619)--(3.512,4.640)--(3.372,4.677)--(3.232,4.410)%
  --(3.093,4.775)--(2.953,4.455)--(2.813,4.567)--(2.673,4.616)--(2.534,4.718)--(2.394,4.818)%
  --(2.254,4.310)--(2.114,4.175)--(1.975,4.125)--(1.835,4.182)--(1.695,4.275)--(1.555,4.351)%
  --(1.416,4.415);
\gpcolor{color=gp lt color border}
\node[gp node right] at (8.936,3.443) {insertion time};
\gpcolor{rgb color={0.867,0.094,0.122}}
\draw[gp path] (9.120,3.443)--(10.036,3.443);
\draw[gp path] (10.500,2.317)--(10.360,2.232)--(10.220,2.257)--(10.080,2.268)--(9.941,2.258)%
  --(9.801,2.245)--(9.661,2.202)--(9.521,2.186)--(9.382,2.180)--(9.242,2.160)--(9.102,2.262)%
  --(8.962,2.228)--(8.823,2.186)--(8.683,2.233)--(8.543,2.169)--(8.403,2.163)--(8.264,2.130)%
  --(8.124,2.110)--(7.984,2.124)--(7.844,2.081)--(7.705,2.075)--(7.565,2.021)--(7.425,1.985)%
  --(7.285,1.907)--(7.146,1.939)--(7.006,1.966)--(6.866,1.952)--(6.726,1.945)--(6.587,1.994)%
  --(6.447,2.003)--(6.307,2.011)--(6.167,2.067)--(6.028,2.091)--(5.888,2.087)--(5.748,2.099)%
  --(5.608,2.076)--(5.468,2.079)--(5.329,2.128)--(5.189,2.113)--(5.049,2.188)--(4.909,2.232)%
  --(4.770,2.236)--(4.630,2.267)--(4.490,2.295)--(4.350,2.375)--(4.211,2.329)--(4.071,2.369)%
  --(3.931,2.480)--(3.791,2.537)--(3.652,2.597)--(3.512,2.586)--(3.372,2.656)--(3.232,2.747)%
  --(3.093,2.943)--(2.953,3.002)--(2.813,3.061)--(2.673,3.171)--(2.534,3.322)--(2.394,3.551)%
  --(2.254,3.889)--(2.114,3.983)--(1.975,4.145)--(1.835,4.480)--(1.695,4.979)--(1.555,5.763)%
  --(1.416,7.130);
\gpcolor{color=gp lt color border}
\gpsetlinetype{gp lt border}
\gpsetlinewidth{1.00}
\draw[gp path] (1.136,7.251)--(1.136,0.985)--(10.919,0.985)--(10.919,7.251)--cycle;
\gpdefrectangularnode{gp plot 1}{\pgfpoint{1.136cm}{0.985cm}}{\pgfpoint{10.919cm}{7.251cm}}
\end{tikzpicture}
}
  \caption{The trade-off between search time and insertion time as a function of $\eps$.}
   \figlabel{epsilon}
\end{figure}

\subsubsection{The Race}

Next, we tested the performance of todolists against a number of other
common dictionary data structures, including skiplists, red-black trees,
scapegoat trees, and treaps. As a baseline, we also measured the search
performance of two static data structures: sorted arrays, and perfectly
balanced binary search trees.

In these tests, the value of $n$ varied from $25,000$ to $2\times 10^6$
in increments of $25,000$.   Each individual test followed the same
pattern as in the previous section and consisted of $n$ insertions
followed by $5n$ searches.

\paragraph{Searches: Todolists win.}
The timing results for the searches are shown in \figref{bigsearch}. In
terms of search times, todolists---with $\eps = 0.2$ and $\eps=0.35$---are
the winners among all the dynamic data structures, and even match the
performance of statically-built perfectly-balanced binary search trees.
The next fastest dynamic data structures are red-black trees which,
for larger $n$ have a search time of roughly 1.4 times that of
perfectly-balanced binary search trees.

Surprisingly, todolists beat red-black trees because of their memory
layout, not because they reduce the number of comparisons. In
these experiments, the average number of comparisons done by
red-black trees during a search was measured to be $\alpha\log n$
for $\alpha\in[1.02,1.03]$.\footnote{This incredibly good performance
of red-black trees created by inserting random data has been observed
already by Sedgewick \cite{sedgewick:left-leaning}, who conjectures
that the average depth of a node in a such a red-black tree is $\log
n-1/2$.} This is substantially less than the number of comparisons done
by todolists, which is about $1.2\log n$ (for $\eps=0.2$) and $1.35\log
n$ for ($\eps=0.35$). The optimal binary search trees also have a
similarly efficient memory layout because the algorithm that constructs
them allocates nodes in the order they are encountered in a pre-order
traversal. A consequence of this is that the left child of any node,
$u$, is typically placed in a memory location adjacent to $u$.  Thus,
during a random search, roughly half the steps proceed to an adjacent
memory location.

\begin{figure}
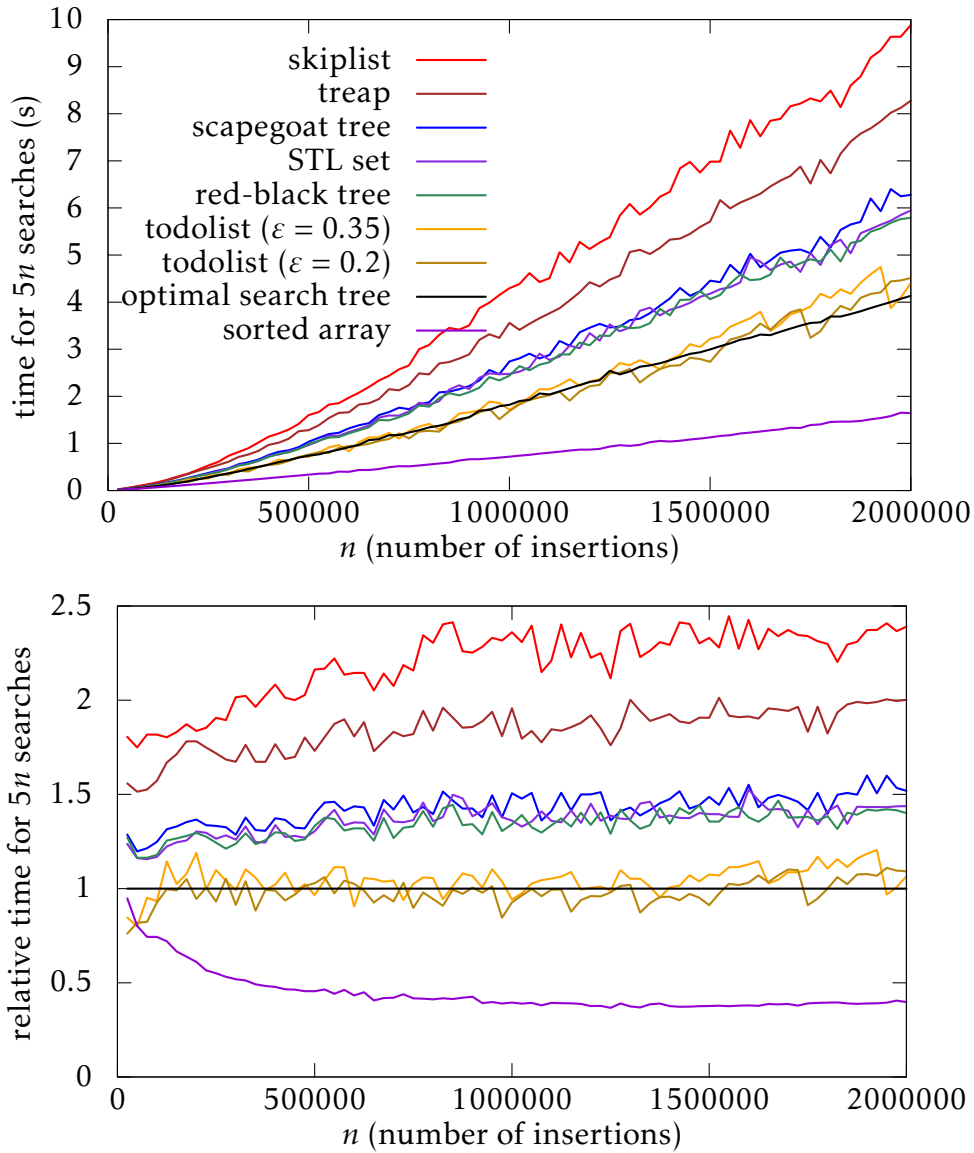

  \centering{\input{experiments/bigtest-find.tex}}
  \centering{\input{experiments/bigtest-find-norm.tex}}
  \caption{Search time comparison between different data structures. The top graph shows absolute times, in seconds. The bottom graph shows relative times, normalized by the time taken in the optimal search tree.}
   \figlabel{bigsearch}
\end{figure}

\paragraph{Insertions: Todolists lose.}
The timing results for the insertions are shown in
\figref{biginsertion}. This is where the other shoe drops.  Even with
$\eps=0.35$, insertions take about three to four times as long in
a todolist as in a red-black tree.  Profiling the code shows that
approximately 65\% of this time is spent doing partial rebuilding and
another 6\% is due to global rebuilding.

One perhaps surprising result is that scapegoat trees, which are also
based on partial rebuilding, outperform todolists in terms of insertions.
This is because scapegoat trees are opportunistic, and only perform
partial rebuilding when needed to maintain a small height.  Randomly built
binary search trees have logarithmic depth, so scapegoat trees do very
little rebuilding in our tests.  In a similar test that inserts elements
in increasing order, scapegoat tree insertions took approximately 50\%
longer than todolist insertions.

\begin{figure}
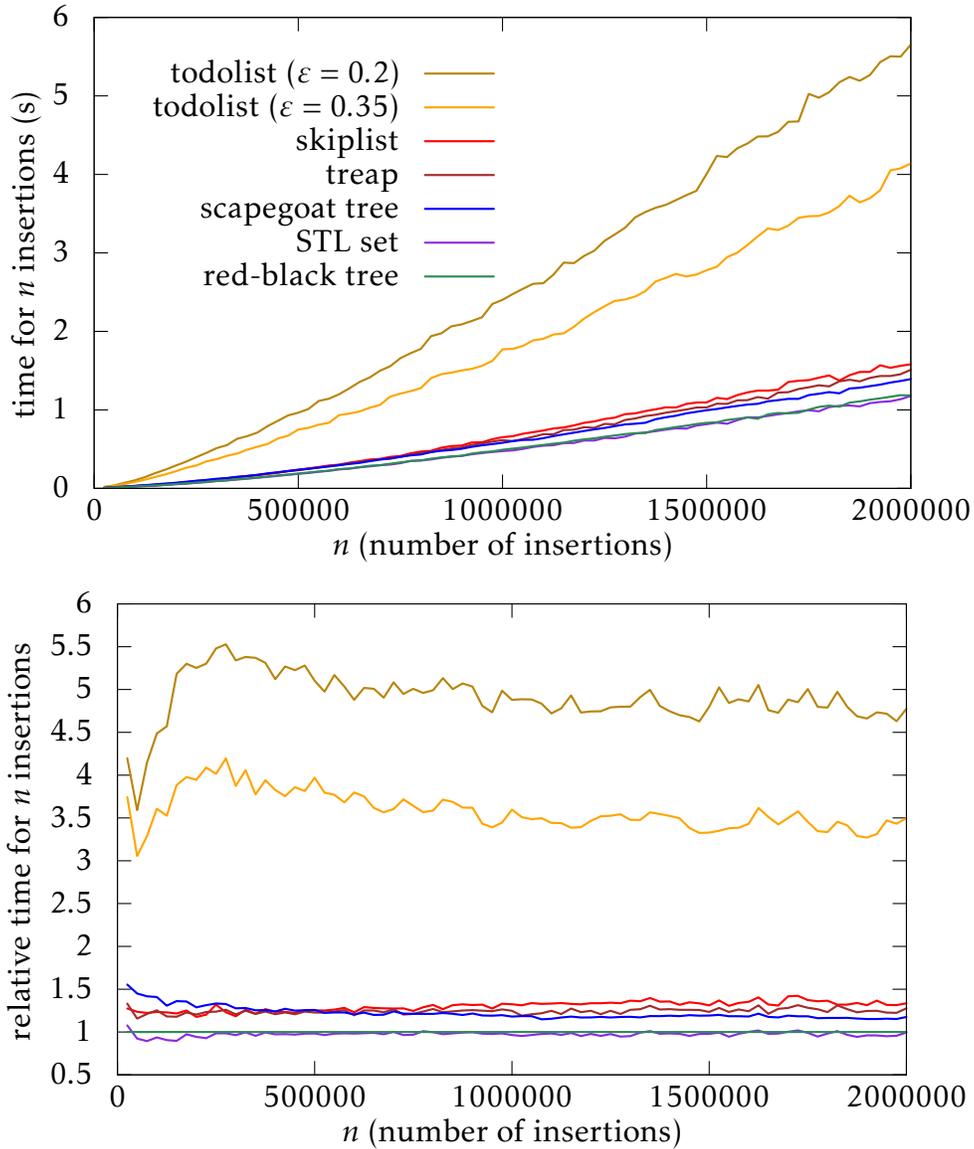

  \centering{\input{experiments/bigtest-add.tex}}
  \centering{\input{experiments/bigtest-add-norm.tex}}
  \caption{Insertion time comparison between different data structures. The top graph shows absolute times, in seconds. The bottom graph shows relative times, normalized by the running time of the red-black tree.}
   \figlabel{biginsertion}
\end{figure}

\section{Conclusion}

If searches are much more frequent than updates, then todolists may
be the right data structure to use.  When implemented properly, their
search times are difficult to beat.  They perform $\log_{2-\eps}
n$ comparisons and roughly half these lead to an adjacent array
location. Thus, a search in a todolist should incurs only about
$\frac{1}{2}\log_{2-\eps} n$ cache misses on average.   $B$-trees
\cite{bayer.mccreight:organization} and their cache-oblivious counterparts
\cite{bender.demaine.ea:cache-oblivious,bender.duan.ea:locality-preserving}
can reduce this to $O(\log_B n)$, where $B$ is the size of a cache line,
but they have considerably higher implementation complexity and running-time
constants.

On the other hand, todolists leave a lot to be desired in terms of
insertion and deletion time.  Like other structures that use partial
rebuilding, the restructuring done during an insertion takes time
$\Omega(\log n)$, so is non-negligible.  The implementation of the
insertion algorithm used in our experiments is fairly naïve and could
probably be improved, but it seems unlikely that its performance will
ever match that of, for example, red-black trees.

\section*{Acknowledgement}

The authors are grateful to Rolf~Fagerberg for helpful discussions
and suggestions.


\bibliographystyle{plain}
\bibliography{todolist}

\end{document}